\documentstyle[11pt]{article}

\begin{document}
\begin{titlepage}

\begin{flushright}
QMW--PH--94--14\\
{\bf hep-th/9501094}\\
January $21^{st}$, 1995
\end{flushright}

\begin{center}

\baselineskip25pt
{\LARGE {\bf TIME-SYMMETRIC INITIAL DATA SETS IN 4--D DILATON GRAVITY}}

\vspace{1cm}

{\large {\bf Tom\'as Ort\'{\i}n}
\footnote{E--mail address: {\tt t.ortin@qmw.ac.uk}}\\
{\it Department of Physics}\\
{\it Queen Mary and Westfield College}\\
{\it Mile End Road, London E1 4NS, U.K.}}

\end{center}

\vspace{1.5cm}


\begin{abstract}

I study the time--symmetric initial--data problem in theories with a
massless scalar field (dilaton), free or coupled to a Maxwell field in
the stringy way, finding different initial--data sets describing an
arbitrary number of black holes with arbitrary masses, charges and
asymptotic value of the dilaton.

The presence of the scalar field gives rise to a number of interesting
effects.  The mass and charges of a single black hole are different in
its two asymptotically flat regions across the Einstein--Rosen bridge.
The same happens to the value of the dilaton at infinity.  This forbids
the identification of these asymptotic regions in order to build
(Misner) wormholes in the most naive way.  Using different techniques, I
find regular initial data for stringy wormholes.  The price payed is the
existence singularities in the dilaton field.  The presence of a
single--valued scalar seems to constrain strongly the allowed topologies
of the initial space--like surface.  Other kinds of scalar fields
(taking values on a circle or being defined up to an additive constant)
are also briefly considered.

\end{abstract}

\end{titlepage}

\newpage

\baselineskip14pt
\pagestyle{plain}


\section*{Introduction}

The usual procedure of getting exact solutions in General Relativity
({\it i.e.} imposing some symmetries on the solutions, substituting an
appropriate ansatz and solving the differential equations) has an
important drawback: one doesn't know what physical system a solution is
going to describe till one actually gets it.  In the simplest cases one
can expect a black hole solution, a cosmological solution etc., but in
more complex cases, following that recipe, one might never find a
solution describing the evolution of the system one is interested in.

If one wants to decribe the time--evolution of a system consisting, for
example, of two black holes subjected to their mutual attraction, the
initial--value formulation of the Einstein equations\footnote{See, for
instance, Ref.~\cite{kn:Wald} for a comprehensive presentation whose
main lines I will follow in the first Section.} is far more appropriate.
Obtaining exact (complete) solutions is still probably hopeless but, at
least, if one has the right initial data, one knows which system one is
working with and one knows that there exists such a solution.  Then it
makes sense to use numerical methods to evolve the initial data.  Many
interesting results have been obtained in this way.

The problem of finding the {\it right} initial data remains, but it is a
much more tractable one.  In General Relativity (as in any other theory
with gauge freedom) the initial data cannot be chosen arbitrarily but
have to satisfy certain constraints.  Solving these constraints is what
is called the {\it initial--data problem}.  Solving the initial--data
problem is interesting not only to get something whose evolution can be
studied.  An initial--data set contains a great deal of information
about the system since it already solves part of the Einstein equations
(the constraints).

My goal in this paper is to find initial--data sets which solve a
simple case of the initial--data problem: the time--symmetric case
\cite{kn:MisWhe,kn:Bri}. The solutions to the time--symmetric
initial--data problem will describe an arbitrary number of non--rotating
black holes which are momentarily at rest, that is, at the moment at
which they ``bounce".  I will also look for (Misner) wormhole initial
data.

Solutions to this problem are already known for the vacuum and
electrovacuum case \cite{kn:MisWhe,kn:BriLin,kn:Lin,kn:Mis}.  Here I
will work with two different theories with a scalar field (dilaton):
Einstein plus a dilaton (also called Einsein-Higgs system in the
literature) and Einstein--Maxwell plus a dilaton which couples to the
Maxwell field in the stringy way.  In this article I present solutions
analogous to those in Refs.~\cite{kn:MisWhe,kn:BriLin,kn:Lin} for the
former cases, with a number of peculiar features.  Perhaps the most
important one is that in most of these solutions the dilaton field has
different asymptotic values in different asymptotic regions.  Since the
zero mode of the dilaton is physically meaningful (in string theory it
is the coupling constant) those regions are physically different
universes.

Another consequence of the existence of different asymptotic values of
the dilaton arises when one tries to find wormhole initial data.
Naively, one would build a wormhole by identifying two different
asymptotic regions linked by an Einstein--Rosen bridge \cite{kn:EinRos}
and the fields defined in them.  In this case one would get a
multivalued dilaton field which is unacceptable unless, by some reason,
it is assumed that the scalar field takes values in a circle or its
zero--mode has no physical meaning, it is ``pure gauge".  Solutions of
this kind will be presented in this paper for the
Einstein--Maxwell--dilaton and for the Einstein--dilaton cases.
Although it is extremely hard to find a wormhole solution for the
Einstein--Maxwell--dilaton system with a single--valued dilaton, one
solution of this kind will also be presented.  This solution is not
smooth.  The string--frame metric is smooth (actually, it is exactly
equal to the metric of a Reissner--Nordstr\"om wormhole in the Einstein
frame) but the dilaton field still has many singularities.

One ``experimentally" observes that there is no way to build wormhole
initial data without introducing at the same time unwanted singularities
in the metric, in the dilaton field, or in both; or without assuming
that the dilaton field lives in a circle or that it is defined up to a
constant.  One can consider other kinds of scalar fields, different from
the string theory dilaton field, with those properties.  For them,
finding wormhole solutions is easier and the resulting configurations
are very interesting.  In the last Section I will discuss the role a
single--valued scalar field seems to play in limiting the possible
topologies of the initial Cauchy surface.

What can be learnt from all these solutions?  First of all, the mere
existence of some of them is interesting {\it per se}, as I am going
to explain.  Secondly, a number of issues can be investigated using
them: no--hair theorems, area theorems, cosmic censorship (first
studied in Ref.~\cite{kn:Gibb}), critical behavior in the gravitational
collapse \cite{kn:critnumeric,kn:critanalytic} as analyzed
in Ref.~\cite{kn:Woj} etc.

The purely scalar case has a special interest since no static
black--hole solutions with a non--trivial scalar field exist.  Different
no--hair theorems forbid its existence \footnote{The closest to a black
hole with scalar hair is the solution given in Ref.~\cite{kn:SBH}.  For
references on no--hair theorems see, for instance, the recent paper
Ref.~\cite{kn:nohair}}.  If solutions to the initial--data problem
describing many black holes with non--trivial scalar hair exist, two
processes must take place: first, all their scalar hair must be radiated
away to infinity, and second, the black holes must merge.  The endpoint
of these processes should be a single black hole of the Kerr family,
according to the Carter--Israel conjecture, and, in our case in which
there is no angular momentum, a Schwarzschild black hole.  The area of
its event horizon must be bigger than the sum of those of the black
holes one started with, and its mass must be smaller.  Then one can
compare initial and final states and give bounds on the energy radiated
away by the system \cite{kn:Gibb}.  This provides a strong test of many
ideas widely believed to hold in General Relativity.  I will
not investigate these issues in this paper, though, and I will limit
myself to the search and indentification of the seeked for
initial--data leaving that investigation for further publications.

In the usual Einstein--Maxwell plus dilaton case in which the dilaton is
not coupled to the Maxwell field the no--hair theorems state that the
only black--hole--type solution is the Reissner--Nordstr\"om solution,
with trivial (constant, zero charge) dilaton.  However, when the dilaton
couples to the Maxwell field (that is, in the low energy string theory
case), a non--trivial dilaton field whose charge is determined by the
electric charge is required in order to get the dilaton black holes of
Refs.~\cite{kn:dilbh}.  These are the only black--hole--type solutions
of this theory according to the unicity theorem of
Ref.~\cite{kn:Masood}.  In some sense the situation os analogous to that
of the purely scalar case: there is a family of solutions for which the
dilaton charge is a free parameter (found in Ref.~\cite{kn:SBH} for the
purely scalar case and in Refs.~\cite{kn:SBH2,kn:BMQ} for this case) but
only for a determined value of the dilaton charge the solution is not
singular and describes a black hole (zero for the purely scalar case and
$-e^{-2\phi_{0}}Q^{2}/2M$ for this case).

It is reasonable to expect that the endpoint of a non--rotating
low--energy string theory black hole will be a static dilaton black hole
and that the dilaton charge will evolve (increasing or decreasing) in
such a way that, in the end, it will have the {\it right} value, no
singularities will be present and cosmic censorship will be enforced.
In this process the area must not decrease and the mass must not
increase.  Observe that, if all the scalar hair disppears, as one
expects to heppen in the Einstein--dilaton case, the endpoint would
exhibit naked singularities.  Why the dilaton behaves so differently and
if whether it really does or not (violating cosmic censorship) are two
important questions that can be addressed in this framework.  The
initial--data sets that I will present are most suited for investigating
these issues.

The structure of this article is the following:

In Section~\ref{sec-inidata} I set up the time--symmetric the
initial--data problem for the theories considered in this paper, define
the charges and asymptotic values of the fields and explain some
conventions.

In Section~\ref{sec-ansatzs} I present the ansatzs, substitute them into
the constraints and reduce them to differential equations with
well--known solutions.

In Section~\ref{sec-BH} I solve those equations to find black hole
solutions. A technique for generating solutions of the Einstein--Maxwell
plus dilaton initial--data problem starting from solutions of the
Einstein--Maxwell initial data is developed in order to find the initial
data of the known charged dilaton black holes \cite{kn:dilbh}.

In Section~\ref{sec-sphesymm} I analyze the spherically symmetric
black--hole solutions found in the previous section.

In Section~\ref{sec-worm} I solve the equations found in
Section~\ref{sec-ansatzs} to find wormhole initial--data.  This turns
out to be a complicated problem and it will be necessary to use the
solution--generating technique developed in Section~\ref{sec-BH} to
solve it, although the result will not be completely satisfactory for
the reason mentioned above.

In Section~\ref{sec-conclussions} I discuss the results and present the
conclussions.

\newpage


\section{The time--symmetric initial data problem}
\label{sec-inidata}

A set of initial data for a space--time $({\cal M},g_{\mu\nu})$ in
General Relativity consists of
\begin{enumerate}
\item A spacelike hypersurface $\Sigma$ determined
      by its normal unit\footnote{We follow the conventions
      in Ref.~\cite{kn:KLOPP}. In particular the signature is $(+---)$,
      although I will use the signature $(+++)$ for the intrinsic
      three--metric on $\Sigma$. Greek indices go from $0$ to $3$ and
      Latin indices from  $1$ to $3$. The dual of the Maxwell tensor is
      ${}^{\star}F^{\mu\nu} =\frac{1}{2\sqrt{-g}}
      \epsilon^{\mu\nu\rho\sigma}
      F_{\rho\sigma}$ with $\epsilon^{\hat{0}\hat{1}\hat{2}\hat{3}}
      =+i$.} vector $n^{\mu}$, $n^{2}=+1$,
\item The induced metric on $\Sigma$, $h_{\mu\nu} =g_{\mu\nu}
      -n_{\mu}n_{\nu}$,
\item The extrinsic curvature  $K_{\mu\nu}=
      \frac{1}{2}{\cal L}_{n}h_{\mu\nu}$ of $\Sigma$.
\end{enumerate}

If, in addition to $g_{\mu\nu}$, there are more fundamental fields, one
has to consider also their initial data.  Both $h_{\mu\nu}$ and
$K_{\mu\nu}$ can be expressed as the intrinsic metric
${}^{(3)}g_{\hat{\imath}\hat{\jmath}}$ and tensor
${}^{(3)}K_{\hat{\imath}\hat{\jmath}}$.  It is also convenient to use
the covariant derivative $D_{\mu}$ associated to $h_{\mu\nu}$,
\begin{equation}
D_{\mu} T_{\nu_{1}\ldots\nu_{n}}{}^{\rho_{1}\ldots\rho_{m}} \equiv
h_{\mu}{}^{\alpha} h_{\nu_{1}}{}^{\beta_{1}} \ldots
h_{\nu_{n}}{}^{\beta_{n}} h^{\rho_{1}}{}_{\gamma_{1}} \ldots
h^{\rho_{m}}{}_{\gamma_{m}} \nabla_{\alpha}
T_{\beta_{1}\ldots\beta_{n}}{}^{\gamma_{1}\ldots\gamma_{m}}\, ,
\end{equation}
which is equivalent to the covariant derivative
${}^{(3)}\nabla_{\hat{\imath}}$ associated to the intrinsic metric
${}^{(3)}g_{\hat{\imath}\hat{\jmath}}$.

The projections $n^{\mu}(G_{\mu\nu}-T_{\mu\nu})=0$ of the Einstein
equations contain only first time--derivatives of the metric.
Therefore, they are not dynamical equations, but constraints on the
metric and its first time--derivatives which have to be satisfied, in
particular, by the initial data on $\Sigma$.  There are two sets of
constraints: $n^{\mu}n^{\nu} (G_{\mu\nu}-T_{\mu\nu})=0$ and $n^{\mu}
h_{\rho}{}^{\nu} (G_{\mu\nu} -T_{\mu\nu})=0$.  Both result into
equations for $h$ and $K$ or their intrinsic counterparts.

The reason for the existence of these constraints is the gauge freedom
of General Relativity.  If there are more fields with gauge freedom some
projections of their equations of motion will typically be constraints
for their initial data, and the equations for $h$ and $K$ will have to
be supplemented with them.

In the special case in which $\Sigma$ is a surface of time--symmetry
(i.e. invariant under time--reflection with respect to $\Sigma$ itself
\cite{kn:MisWhe}) $K$ vanishes and the equations take a very simple form
\begin{eqnarray}
{}^{(3)}R  -2n^{\mu}n^{\nu}T_{\mu\nu} & = & 0\, ,
\nonumber \\
& &
\nonumber \\
n^{\mu}h_{\rho}{}^{\nu}T_{\mu\nu} & = & 0\, .
\label{eq:constEins}
\end{eqnarray}

All the theories I want to consider can be described by different
truncations of the following action \cite{kn:KLOPP}

\begin{equation}\label{eq:action}
S  = \frac{1}{16\pi} \int d^{4}x \sqrt{-g} \biggl \{-R
+2(\partial\phi)^{2} -e^{-2\phi} F^{2} \biggr \}\, ,
\end{equation}

\noindent which is itself a truncation of the low--energy string theory
effective action in four dimensions written in the Einstein frame.  I
will consider different cases: the full theory, the Einstein--Maxwell
case by setting $\phi=0$ everywhere and ignoring its equation of motion,
the purely scalar case by setting $F_{\mu\nu}=0$ and ignoring its
equations of motion, and, in the obvious way, the vacuum case. I will
not consider the Einstein--Maxwell plus uncoupled dilaton case.

In string theory, the dilaton $\phi$ is a scalar field that takes values
in $R$.  This is the kind of scalar field I will work with, unless I
explicitly indicate otherwise.  However, in some situations, it will be
interesting to consider different scalar fields, taking values in
$S^{1}$ or defined up to an additive constant.

Sometimes I will also be interested in the string--frame metric

\begin{equation}
g^{string}_{\mu\nu} =e^{2\phi}g_{\mu\nu}\, ,
\end{equation}

\noindent or other metrics that can be obtained by rescaling the
Einstein-frame metric with certein powers of $e^{\phi}$.

A complete initial--data set for the theory described by the action
Eq.~(\ref{eq:action}) consists of the initial Cauchy surface $\Sigma$,
its induced metric $h_{\mu\nu}$ and its extrinsic curvature
$K_{\mu\nu}$, the values of $\phi$ and $n^{\mu}\nabla_{\mu}\phi$ on
$\Sigma$ and the electric and magnetic fields on $\Sigma$, defined by
\begin{equation}
E_{\mu} = n^{\nu}F_{\nu\mu}\, ,
\hspace{1cm}
B_{\mu} = -in^{\nu}\, {}^{\star}F_{\nu\mu}\, .
\end{equation}
If $\Sigma$ is a surface of time--symmetry, then
\begin{equation}\label{eq:symdil}
n^{\mu} \nabla_{\mu} \phi =0\, ,
\end{equation}
on $\Sigma$.

Now, what it is needed to study the initial--data problem of the theory
given by the action Eq.~(\ref{eq:action}) is
\begin{description}
\item[(i)] The energy--momentum
tensor, which is most conveniently written as follows

\begin{equation}
T_{\mu\nu} =
e^{-2\phi}[F_{\mu\rho}F_{\nu}{}^{\rho}
-{}^{\star}F_{\mu\rho} {}^{\star}F_{\nu}{}^{\rho}]
-2[\nabla_{\mu}\phi\nabla_{\nu}\phi
-{\textstyle\frac{1}{2}}g_{\mu\nu}(\nabla\phi)^{2}]\, ,
\end{equation}

\noindent and which has to be substituted in Eqs.~(\ref{eq:constEins})
and

\item[(ii)] The equations of motion of $\phi$ and $F_{\mu\nu}$ to find
the possible constraints contained in them.  These are

\begin{eqnarray}
\nabla^{2}\phi -{\textstyle\frac{1}{2}}e^{-2\phi}F^{2} & = & 0\, ,
\label{eq:dilaton}
\\
& &
\nonumber \\
\nabla_{\mu}(e^{-2\phi}F^{\mu\nu}) & = & 0\, ,
\label{eq:Maxwell}
\\
& &
\nonumber \\
\nabla_{\mu}{}^{\star}F^{\mu\nu} & = & 0\, .
\label{eq:Bianchi}
\end{eqnarray}

\end{description}

The equation of motion of the dilaton, Eq.~(\ref{eq:dilaton}), leads to
no contraint on $\Sigma$.  This was expected since there is no gauge
invariance associated to $\phi$, which is a fundamental physical
field\footnote{It should be stressed that the zero mode of the dilaton
is physically meaningful.  The symmetry $\phi\rightarrow\phi^{\prime}
=\phi +constant$ of the action Eq.~(\ref{eq:action}) means that given a
solution of its equations of motion, another solution can be obtained by
shifting the value of the dilaton zero--mode.  However this is not a
gauge symmetry and both solutions have to be regarded as physically
inequivalent.  Exactly the opposite happens to the electrostatic
potential which is not a physical field.  Shifting the value of its zero
mode is not just a symmetry, but it is part of a gauge symmetry.  The
zero mode of the electrostatic potential is physically meaningless.
This will be important later, when trying to build wormhole solutions.
On the other hand, for the same reasons, the dilaton charge is not a
conserved charge while the electric charge is.  In spite of this fact,
the dilaton charge is a useful parameter that I will use to describe the
solutions found.}.  We have included the Bianchi identity of $F$
Eq.~(\ref{eq:Bianchi}) because, although it is not an equation of motion
of the vector field $A$, it constrains $F$ on the initial surface
$\Sigma$.  These constraints are

\begin{eqnarray}
n^{\mu}\nabla^{\nu}(e^{-2\phi}F_{\mu\nu}) & = & 0\, ,
\nonumber \\
& &
\nonumber \\
n^{\mu}\nabla^{\nu}{}^{\star}F_{\mu\nu} & = & 0\, .
\label{eq:constMax}
\end{eqnarray}

The complete set of equations that have to be solved consists of
Eqs.~(\ref{eq:constEins}) plus Eqs.~(\ref{eq:constMax}), taking into
account Eq.~(\ref{eq:symdil}).  It is convenient to express them in
terms of the instrinsic geometric objects of $\Sigma$.  One gets the
following set of equations

\begin{eqnarray}
{}^{(3)}R -2\, {}^{(3)}g^{\hat{\imath}\hat{\jmath}}
e^{-2\phi} (E_{\hat{\imath}} E_{\hat{\jmath}}
+B_{\hat{\imath}} B_{\hat{\jmath}})
-2\, {}^{(3)}g^{\hat{\imath}\hat{\jmath}}
{}^{(3)}\nabla_{\hat{\imath}} \phi {}^{(3)}\nabla_{\hat{\jmath}}
\phi & = & 0\, ,
\nonumber \\
& &
\nonumber \\
{}^{(3)}\nabla_{\hat{\imath}} ({}^{(3)}g^{\hat{\imath}\hat{\jmath}}
e^{-2\phi} E_{\hat{\jmath}}) & = & 0\, ,
\nonumber \\
& &
\nonumber \\
{}^{(3)}\nabla_{\hat{\imath}} ({}^{(3)}g^{\hat{\imath}\hat{\jmath}}
B_{\hat{\jmath}}) & = & 0\, ,
\label{eq:ini}
\end{eqnarray}

\noindent which define the time---symmetric initial--data problem for the
above theory.  In what follows I will omit the indices ${}^{(3)}$ in
most places for simplicity.

The kind of solutions of these equations that I am looking for are
asymptotically flat and are determined by the mass ($M$), the electric
magnetic charges and dilaton charges ($Q$, $P$ and $\Sigma$,
respectively) and the asymptotic value of the dilaton at infinity
$\phi_{\infty}$.  They are defined in the limit
$r=|\vec{x}|\rightarrow\infty$ as follows:

\begin{eqnarray}
{}^{(3)}g_{rr} & \sim & 1 + \frac{2M}{r}\, ,
\nonumber \\
& &
\nonumber \\
E_{\hat{\imath}} & \sim & Q\,\frac{x_{\hat{\imath}}}{r^{3}}\, ,
\nonumber \\
& &
\nonumber \\
B_{\hat{\imath}} & \sim & P\, \frac{x_{\hat{\imath}}}{r^{3}}\, ,
\nonumber \\
& &
\nonumber \\
\phi & \sim & \phi_{\infty} +\frac{\Sigma}{r}\, ,
\end{eqnarray}

\noindent so they coincide with those of the known exact static
solutions \cite{kn:dilbh,kn:KLOPP}. When there is more than one
asymptotically flat region, one has to indentify first the
coordinate that plays the same role as $r$ in that region and then one
can define as before the charges that observers in that region would
measure. As we will see they are different in general.

In the following sections I am going to find different families of
solutions to Eqs.~(\ref{eq:ini}) describing black holes and wormholes.


\section{The ansatzs}
\label{sec-ansatzs}

To find solutions to the equations of the previous section I am going
to use a combination of ``tricks" previously used in the literature.
First I make the following ansatz for the three--metric $dl^{2}
={}^{(3)}g_{\hat{\imath}\hat{\jmath}} dx^{\hat{\imath}}
dx^{\hat{\jmath}}$

\begin{equation}
dl^{2}=W(\vec{x})dl_{\xi}^{2}\, ,
\label{eq:ansatz}
\end{equation}

\noindent where $dl^{2}_{\xi}$ is another three--metric which, later, I
will choose to be either the flat Euclidean metric

\begin{equation}
dl_{\flat}^{2}=d\vec{x}^{2}\, ,
\label{eq:flat}
\end{equation}
or the metric of an $S^{1}\times S^{2}$ ``doughnut"
\begin{equation}
dl_{D}^{2}=d\mu^{2}+d\theta^{2}+\sin^{2}\theta d\phi^{2}\, .
\label{eq:doughnut}
\end{equation}

In terms of the curvature and covariant derivative of the $\xi$--metric
the curvature of the three--metric I want to find is given by

\begin{equation}
{}^{(3)}R=W^{-1} \biggl [ {}^{(3)}R_{\xi}-2\nabla_{\xi}^{2}\log W-
{\textstyle\frac{1}{2}}(\nabla_{\xi}\log W)^{2} \biggr ] \, .
\end{equation}

\noindent The value of ${}^{(3)}R_{\xi}$ in the two cases I am
interested in is ${}^{(3)}R_{\flat}=0$ and ${}^{(3)}R_{D}=2$.

I am going to restrict myself to the purely electric case
$B_{\hat{\imath}}=0$. The purely magnetic one can be obtained through
the duality transformation

\begin{eqnarray}
e^{-2\phi}E_{\hat{\imath}} & \rightarrow & B_{\hat{\imath}}\, ,
\nonumber \\
& &
\nonumber \\
B_{\hat{\imath}} & \rightarrow & -e^{-2\phi}E_{\hat{\imath}}\, ,
\nonumber \\
& &
\nonumber \\
\phi & \rightarrow & -\phi\, .
\label{eq:dual}
\end{eqnarray}

In general it does not make sense to perform a continuous duality
rotation since the truncation of $N=4$, $d=4$ supergravity
Eq.~(\ref{eq:action}) (in which the axion field is absent) would not be
consistent.

It is also easier to work with the electrostatic potential $Z$ defined
by

\begin{equation}
E_{\hat{\imath}}=-\partial_{\hat{\imath}}Z\, .
\end{equation}

\noindent Substituting all this into Eqs.~(\ref{eq:ini}) one arrives at

\begin{eqnarray}
\nabla_{\xi}^{2} \log W +{\textstyle\frac{1}{4}} (\nabla_{\xi}
\log W)^{2} +e^{-2\phi}(\nabla_{\xi}Z)^{2} +(\nabla_{\xi}\phi)^{2}-
{\textstyle\frac{1}{2}} {}^{(3)}R_{\xi} & = & 0\, ,
\label{eq:ini1}
\\
& &
\nonumber \\
\nabla_{\xi}(W^{\frac{1}{2}}e^{-2\phi}\nabla_{\xi}Z) & = & 0\, .
\label{eq:ini2}
\end{eqnarray}

It is impossible to give a more specific unique ansatz for the different
situations I am going to consider here.  I will make ansatzs case by
case, including also the well--known vacuum and electrovacuum cases for
completeness.


\subsection{Vacuum}

The equation that one has to solve is Eq.~(\ref{eq:ini1}) with
$Z=\phi=0$

\begin{equation}
\nabla_{\xi}^{2}\log W +{\textstyle\frac{1}{4}} (\nabla_{\xi} \log
W)^{2} - {\textstyle\frac{1}{2}} {}^{(3)}R_{\xi} = 0\, .
\end{equation}

Now a further ansatz can be made: the function $W$ is a power of a
function $\chi$ such that there are no terms proportional to the square
of the derivative of $\chi$ in the above equation.  This implies

\begin{equation}
W  =  \chi^{4}\, ,
\end{equation}

\noindent where $\chi$ satisfies the equation

\begin{equation}
(\nabla_{\xi}^{2} -{\textstyle\frac{1}{8}} {}^{(3)}R_{\xi})\chi
 =  0\, .
\end{equation}

As it is well known, this equation is easy to solve for the two
$\xi$--metrics I am going to consider here
Eqs.~(\ref{eq:flat}),(\ref{eq:doughnut}).


\subsection{Einstein--Maxwell}

The equations that have to be solved are
Eqs.~(\ref{eq:ini1}),(\ref{eq:ini2}) with $\phi=0$

\begin{eqnarray}
\nabla_{\xi}^{2}\log W+ {\textstyle\frac{1}{4}} (\nabla_{\xi} \log
W)^{2} + (\nabla_{\xi}Z)^{2}- {\textstyle\frac{1}{2}} {}^{(3)}R_{\xi} &
= & 0\, ,
\label{eq:inimax1}
\\
& &
\nonumber \\
\nabla_{\xi}(W^{\frac{1}{2}}\nabla_{\xi}Z) & = & 0\, .
\label{eq:inimax2}
\end{eqnarray}

\noindent Again I make a further ansatz: the functions $W$ and $Z$ can
be expressed in terms of two new functions $\chi$ and $\psi$ in the
following way

\begin{eqnarray}
W & = & \psi^{\delta}\chi^{\gamma}\,
\nonumber \\
& &
\nonumber \\
Z & = & \alpha\log\psi + \beta\log\chi\, ,
\end{eqnarray}

\noindent where the contants $\alpha,\beta,\gamma,\delta$ will be
adjusted to cancel crossed terms etc. in the equations. The result is

\begin{eqnarray}
W & = & (\psi\chi)^{2}\, ,
\nonumber \\
& &
\nonumber \\
Z & = & C\pm\log(\psi/\chi)\, ,
\label{eq:emans}
\end{eqnarray}

\noindent where the functions $\psi$ and $\chi$ satisfy

\begin{eqnarray}
(\nabla_{\xi}^{2} -{\textstyle\frac{1}{8}}{}^{(3)}R_{\xi})\chi
& = & 0\, ,
\nonumber \\
& &
\nonumber \\
(\nabla_{\xi}^{2} -{\textstyle\frac{1}{8}}{}^{(3)}R_{\xi})\psi
& = & 0\, .
\label{eq:2harm}
\end{eqnarray}

This is again the same type of equation that was obtained in the vacuum
case.  Here $C$ is an arbitrary integration constant.  Its value can be
changed by a gauge transformation and, thus, solutions with different
values of $C$ are, in fact, physically equivalent.


\subsection{Einstein--dilaton}

The equation that has to be solved is Eq.~(\ref{eq:ini1}) with

\begin{equation}
\nabla_{\xi}^{2}\log W+ {\textstyle\frac{1}{4}} (\nabla_{\xi} \log
W)^{2} + (\nabla_{\xi}\phi)^{2}- {\textstyle\frac{1}{2}} {}^{(3)}R_{\xi}
=  0\, .
\label{eq:inidil}
\end{equation}

This equation is exactly the same as Eq.~(\ref{eq:inimax1}) of the
Einstein--Maxwell initial--data problem with $\phi$ playing now the role
of the electrostatic potential $Z$.  The difference with the
Einstein--Maxwell case is that now there is no constraint analogous to
Eq.~(\ref{eq:inimax2}), and so Eq.~(\ref{eq:inidil}) is the only
constraint on the initial data of this system.  It is safe to try the
same ansatz as in the Einstein Maxwell case:

\begin{eqnarray}
W & = & \psi^{\delta}\chi^{\gamma}\, ,
\nonumber \\
& &
\nonumber \\
\phi & = & \phi_{0}+\alpha\log\psi + \beta\log\chi\, .
\end{eqnarray}

\noindent The result is

\begin{eqnarray}
W & = & (\psi\chi)^{2}(\psi/\chi)^{2a}\, ,
\nonumber \\
& &
\nonumber \\
\phi & = & \phi_{0}\pm\sqrt{1-a^{2}}\log(\psi/\chi)\, ,
\end{eqnarray}

\noindent where the functions $\psi$ and $\chi$ satisfy again
Eqs.~(\ref{eq:2harm}) and the constant $a$ takes values in the interval
$[-1,1]$. $\phi_{0}$ is a constant which will coincide, in general, with
the value of the dilaton at infinity if the functions $\psi$ and $\chi$
are properly normalized.  When there is more than one asymptotically
flat region there is no reason why one should expect the value of the
dilaton at the corresponding infinities to be the same.  In fact, as it
will be shown later on, the asymptotic values of the dilaton are in
general different in different asymptotic regions and $\phi_{0}$ will be
just one of them.

The constant $\phi_{0}$ can be fixed arbitrarily as was the case with
the integration constant $C$ in the Einstein--Maxwell system, i.e. there
is a solution for each $\phi_{0}$ chosen.  However, as it has already
been explained, these solutions are not physically equivalent.

In the Einstein--Maxwell case, the constant $a$ was forced to be equal to
zero by Eq.~(\ref{eq:inimax2}).  This equation is nothing but Gauss'
law, and it enforces the condition of absence of electric charge on the
initial surface $\Sigma$, which it is assumed to be regular everywhere.
The same equation with $\phi$ playing the role of $Z$ would enforce the
absence of dilaton charge on the initial surface.  Thus, for all cases
with $a\neq 0$ it is reasonable to expect the solutions to have net
dilaton charge.  Then they will be very different from all known
solutions in which the charges are either located at a singularity or
the effect of nontrivial topology (Wheeler's ``charge without charge").

As a matter of fact, Eq.~(\ref{eq:inidil}) could be regarded as the
equation for the Einstein--Maxwell plus (unespecified) charged matter
initial--data problem, just by changing $\phi$ by $Z$.  The charge
density would then be given by

\begin{equation}
\rho={}^{(3)}\nabla^{2} Z\, .
\end{equation}

\noindent The analogy is not perfect because this hypothetical matter
should contribute to the density of energy on $\Sigma$,
$n^{\mu}n^{\nu}T_{\mu\nu}$ and no contribution of this kind appears in
Eq.~(\ref{eq:inidil}).


\subsection{Einstein--Maxwell--dilaton}

Now one has to solve the complete Eqs.~(\ref{eq:ini1}),(\ref{eq:ini2}).
To get something different from the previous cases one has to use a
slightly different ansatz\footnote{If we try again with $Z$ proportional
to the logarithms of $\psi$ and $\chi$ one finds that the solutions
have either trivial dilaton or trivial electrostatic potential,
recovering the solutions found on the two latter sections.}:

\begin{eqnarray}
W & = & \psi^{\delta}\chi^{\gamma}\, ,
\nonumber \\
& &
\nonumber \\
\phi & = & \phi_{0}+\epsilon\log\psi + \mu\log\chi\, ,
\nonumber \\
& &
\nonumber \\
Z & = & B \psi^{\alpha}\chi^{\beta}\, .
\end{eqnarray}

\noindent As one might have expected after all the examples studied so
far the result is

\begin{eqnarray}
W & = & (\psi\chi)^{2}(\psi/\chi)^{+2b}\, ,
\nonumber \\
& &
\nonumber \\
e^{-2\phi} & = & e^{-2\phi_{0}}(\psi/\chi)^{-2b}\, ,
\nonumber \\
& &
\nonumber \\
Z & = & C \pm e^{+\phi_{0}}\frac{\sqrt{1-2b^{2}}}{b}(\psi/\chi)^{+b}\, ,
\label{eq:emdans}
\end{eqnarray}

\noindent where $\psi$ and $\chi$ satisfy, yet again,
Eqs.~(\ref{eq:2harm}) and $b$ is a constant that takes values in the
interval $[-1/\sqrt{2},+1/\sqrt{2}]$.

It is possible to find more solutions following a different procedure
that will be explained later in Section~\ref{subsec-emd}.


\section{Black--hole initial data}
\label{sec-BH}

One can get black--hole initial data for all the cases studied in the
previous section by choosing the $\xi$--metric to be the flat Euclidean
metric, which has ${}^{(3)}R_{\flat}=0$.  The three--metric one gets is
conformally flat.  Then, with all the ansatzs made one finds that only
two functions, $\psi$ and $\chi$, harmonic in flat three--space, are
needed to build solutions for all the different cases:

\begin{equation}
\partial_{\hat{\imath}}\partial_{\hat{\imath}}\psi=
\partial_{\hat{\imath}}\partial_{\hat{\imath}}\chi=0\, .
\end{equation}

\noindent These functions will be normalized to $1$ at infinity and will
correspond to point--like sources $\vec{x}_{i}$:

\begin{eqnarray}
\psi & = & 1 + \sum_{i=1}^{N}\frac{\psi_{i}}{|\vec{x}-\vec{x}_{i}|}\, ,
\nonumber \\
& &
\nonumber \\
\chi & = & 1 + \sum_{i=1}^{N}\frac{\chi_{i}}{|\vec{x}-\vec{x}_{i}|}\, .
\label{eq:harmonic}
\end{eqnarray}

For the three--metrics corresponding to these $\psi$ and $\chi$ to be
regular everywhere (except at the points $\vec{x}_{i}$ which have to be
erased from $\Sigma$) the constants $\psi_{i}$ and $\chi_{i}$ have to be
strictly positive.

The $i$th black hole corresponds to an Einstein--Rosen--like bridge
between two asymptotically flat regions or {\it sheets}: the $|\vec{x}|
\rightarrow \infty$ region and the $|\vec{x}-\vec{x}_{i}| \rightarrow 0$
region\footnote{In each case one has to prove first that the limit
$|\vec{x}-\vec{x}_{i}| \rightarrow 0$ indeed corresponds to another
asymptotically flat region.  I will do so later.}.  Therefore $\Sigma$
is a set of $N+1$ sheets, one of them with $N$ holes cut and the
remaining $N$ with a single hole cut and pasted to one of the $N$ holes
of the {\it largest} sheet forming {\it throats} or {\it necks}
(Einstein--Rosen bridges).  In the largest sheet there will be $N$ black
holes.

The corresponding solutions for the vacuum and Einstein--Maxwell cases
were found in Refs.~\cite{kn:MisWhe,kn:BriLin,kn:Lin} and I will
not discuss them here. I will write them down, however, for completeness
and later use.


\subsection{Vacuum}

The solution is given by the metric

\begin{equation}\label{eq:vacbh}
dl^{2} = \chi^{4}d\vec{x}^{2}\, ,
\end{equation}

\noindent with $\chi$ given by Eq.~(\ref{eq:harmonic}) and describes $N$
Schwarzschild ({\it i.e.} no charge nor angular momentum) black holes.
This solution was first found by Misner and Wheeler in
Ref.~\cite{kn:MisWhe}.

For a single black hole, this metric is nothing but the spatial part of
Schwarzschild's in isotropic coordinates.  For more than one black hole
the corresponding exact solution cannot be static and is not known.


\subsection{Einstein--Maxwell}

The solution is given by the metric and electrostatic potential

\begin{eqnarray}
dl^{2} & = & (\psi\chi)^{2} d\vec{x}^{2}\, ,
\nonumber \\
& &
\nonumber \\
Z & = & C \pm \log (\psi/\chi)\, ,
\label{eq:embh}
\end{eqnarray}

\noindent with the functions $\psi$ and $\chi$ given by
Eqs.~(\ref{eq:harmonic}), and describes $N$ electric
Reissner--Nordstr\"om ({\it i.e.} no angular momentum nor scalar hair)
black holes.  This solution was first found by Brill and Lindquist in
Ref.~\cite{kn:BriLin}.

For a single black hole this solution coincides with the spatial part of
the Reissner--Nordstr\"om solution in isotropic coordinates.  If
$\psi=1$ or $\chi=1$, one has the spatial part of the static solution
that describes $N$ {\it extreme} Reissner-Nordstr\"om black holes in
equilibrium.  The remaining situations do not correspond to systems in
static equilibrium and no known solution describes them.


\subsection{Einstein--dilaton}

The metric and dilaton are given by

\begin{eqnarray}
dl^{2} & = & \psi^{2+2a} \chi^{2-2a} d\vec{x}^{2}\, ,
\nonumber \\
& &
\nonumber \\
e^{\phi} & = & e^{\phi_{0}} (\psi/\chi)^{\pm\sqrt{1-a^{2}}}\, ,
\label{eq:edbh}
\end{eqnarray}

\noindent with $\psi$ and $\chi$ given by Eqs.~(\ref{eq:harmonic}).

First of all, observe that by setting $a=1$ or $\psi=\chi$ the vacuum
solutions Eq.~(\ref{eq:vacbh}) are recovered.  If one sets $a=0$ one
recovers the metric of the Einstein--Maxwell solution
Eqs.~(\ref{eq:embh}) and, upon rescaling by
$\exp\{\frac{-2a}{\sqrt{1-a^2}}\phi\}$ one always recovers that metric.

The choice

\begin{eqnarray}
\psi & = & 1-\rho_{0}/2r\, ,
\nonumber \\
& &
\nonumber \\
\chi & = & 1+\rho_{0}/2r\, ,
\nonumber \\
& &
\nonumber \\
a & = & M/\rho_{0}\, ,
\label{eq:esa}
\end{eqnarray}

\noindent with $r=|\vec{x}|$ and $\rho_{0}=\sqrt{M^{2}+\Sigma^{2}}$
corresponds to the surface of time-symmetry of the family of singular
static solutions of Refs.~\cite{kn:SBH}, which are not black holes and
have naked singularities, in agreement with the no--hair theorems of
Refs.~\cite{kn:nohair} and the unicity theorem of Ref.~\cite{kn:Masood}.

Observe that, if one is looking for everywhere regular initial data, one
would choose $\psi = 1+\rho_{0}/2r$ instead of $\psi$ as in
Eq.~(\ref{eq:esa}).  However this does not lead to an exact static
solution of the full set of Einstein's equations.

The solution in Eqs.~(\ref{eq:edbh}) is clearly asymptotically flat in
the limit $|\vec{x}| \rightarrow \infty$ ({\it upper sheet}).  The
metric is also regular everywhere if $|a|<1$ and $\psi>0$ $\chi>0$
verywhere ({\it i.e.} $\psi_{i}>0$, $\chi_{i}>0$ for all $i$).  Now I
want to prove that it is also asymptotically flat in the {\it i}th {\it
lower sheet} $|\vec{x} -\vec{x}_{i}|\equiv r_{i} \rightarrow 0$.  In
this limit the metric looks like this

\begin{equation}\label{eq:asymp}
dl^{2}\sim \biggl (1
+\frac{\psi_{i}^{\prime}}{r_{i}^{\prime}}\biggr )^{2+2a}
\biggl (1 +\frac{\chi_{i}^{\prime}}{r_{i}^{\prime}} \biggr )^{2-2a}
(dr_{i}^{\prime 2} +r_{i}^{\prime 2}d\Omega^{2})\, ,
\end{equation}
where
\begin{eqnarray}
r_{i}^{\prime} & = &
\psi_{i}^{1+a}\chi_{i}^{1-a}r_{i}^{-1}\, ,
\nonumber \\
& &
\nonumber \\
\psi_{i}^{\prime} & = &
\chi_{i}(\psi_{i}/\chi_{i})^{a}
\biggl (1 +\sum_{j\neq i}^{N} \frac{\psi_{i}}{r_{ij}} \biggr )\, ,
\nonumber \\
& &
\nonumber \\
\chi_{i}^{\prime} & = &
\psi_{i}(\psi_{i}/\chi_{i})^{a}
\biggl (1 +\sum_{j\neq i}^{N} \frac{\chi_{i}}{r_{ij}} \biggr )\, ,
\nonumber \\
& &
\nonumber \\
r_{ij} & = & |\vec{x}_{i}-\vec{x}_{j}|\, .
\end{eqnarray}

Equation~(\ref{eq:asymp}) proves that there is one asymptotic region in
each limit $r_{i}\rightarrow 0$ and that this solution describes $N$
black holes represented by their Einstein--Rosen bridges at the moment
of time--symmetry.  The dilaton field is nontrivial and, thus, these
black holes have scalar hair.  There is no exact static solution
describing any of these objects \cite{kn:nohair,kn:SBH}.


\subsection{Einstein--Maxwell--dilaton}
\label{subsec-emd}

This solution is given by

\begin{eqnarray}
dl^{2} & = & \psi^{2+2b}\chi^{2-2b}d\vec{x}^{2}\, ,
\nonumber \\
& &
\nonumber \\
e^{-2\phi} & = & e^{-2\phi_{0}} (\psi/\chi)^{-2b}\, .
\nonumber \\
& &
\nonumber \\
Z & = & C \pm  e^{+\phi_{0}} \frac{\sqrt{1-2b^{2}}}{b}
(\psi/\chi)^{+b}\, .
\label{eq:emdbh1}
\end{eqnarray}

Taking the limit $r_{i}=|\vec{x}-\vec{x}_{i}| \rightarrow 0$ in the
metric one gets Eq.~(\ref{eq:asymp}) with $a$ replaced by $b$, and this
implies again that this solution has $N$ asymptotically flat regions at
the other side of $N$ Einstein--Rosen bridges, so this family of
solutions must indeed describe $N$ charged dilaton black holes.

Amongst this family of solutions I could identify only the case
$b=-1/2,\ \chi=1$ ($W=\psi$) as the spatial part of the exact static
solution describing $N$ extreme electric dilaton black holes in
equilibrium Refs.~\cite{kn:dilbh,kn:KLOPP}

\begin{eqnarray}
ds^{2} & = & \psi^{-1}dt^{2}-\psi d\vec{x}^{2}\, ,
\nonumber \\
& &
\nonumber \\
e^{-2\phi} & = & e^{-2\phi_{0}}\psi\, ,
\nonumber \\
& &
\nonumber \\
F_{\hat{\imath}\hat{0}} & = & \pm
\frac{e^{\phi_{0}}}{\sqrt{2}}\partial_{\hat{\imath}}\psi^{-1}\, .
\end{eqnarray}

The non--extreme dilaton black holes of Refs.~\cite{kn:dilbh,kn:KLOPP}
also possess surfaces of time--symmetry.  However none of them is
described by the initial--data sets of Eqs.~(\ref{eq:emdbh1}).  To find
the family of initial data which describes them one has to follow a
different procedure.

Let $(W,Z)$ be a set of time--symmetric initial data of the
Einstein--Maxwell system.  Then one can build out of it a set of
time--symmetric initial data of the Einstein--Maxwell--dilaton system
$(\tilde{W},\tilde{\phi},\tilde{Z})$ according to the rules

\begin{eqnarray}
\tilde{W} & = &  p(t^{\prime})^{2r}W\, ,
\nonumber \\
& &
\nonumber \\
e^{-2\tilde{\phi}} & = & q(t^{\prime})^{-(1+r)}\, ,
\nonumber \\
& &
\nonumber \\
\tilde{Z} & = & t(Z)\, ,
\end{eqnarray}

\noindent where $p,q,r$ are arbitrary constants and $t(Z)$ is a function
of $Z$ satisfying the differential equation

\begin{equation}\label{eq:differential}
q(t^{\prime})^{-(1+r)}+
2r\frac{t^{\prime\prime\prime}t^{\prime}-
(t^{\prime\prime})^{2}}{(t^{\prime})^{2}}+
\frac{(1+r)^{2}+4r^{2}}{4}
\left(\frac{t^{\prime\prime}}{t^{\prime}}\right)^{2}-1=0\, .
\end{equation}

Here $t^{\prime}$ denotes the derivative of $t(Z)$ with respect to its
argument $Z$.

Finding the most general solution of this equation is a very difficult
problem.  It turns out that there are several choices of the constants
$p,q,r$ which simplify enormously the problem and that those choices
give, too, the solutions I am after.  In particular, for $r=1$ the
solution is simply

\begin{equation}\label{eq:t}
t(Z) = D\left( Ae^{\sqrt{\frac{1-q}{2}}Z} -Be^{-\sqrt{\frac{1-q}{2}}Z}
\right)\, .
\end{equation}

Although it is not obvious, for $q=\frac{1}{2}$, $D=\pm \sqrt{2}
e^{\phi_{0}}$, $p=2e^{-2\phi_{0}}$, $(A+B)^{2}=1$ this is exactly the
transformation that takes the Reissner--Nordstr\"om initial data
Eq.~(\ref{eq:embh}) into the wanted dilaton black hole initial data:

\begin{eqnarray}
dl^{2} & = & \left[ A(\psi/\chi)^{1/2} +B(\psi/\chi)^{-1/2}
\right]^{2}\, ,
\nonumber \\
& &
\nonumber \\
e^{-2\phi} & = & e^{-2\phi_{0}} \left[ A(\psi/\chi)^{1/2}
+B(\psi/\chi)^{-1/2} \right]^{-2}\, ,
\nonumber \\
& &
\nonumber \\
Z & = & C \pm \sqrt{2} e^{\phi_{0}}
\left[ A(\psi/\chi)^{1/2} -B(\psi/\chi)^{-1/2} \right]\, .
\label{eq:emdbh2}
\end{eqnarray}

To see that this is indeed true, these initial data, for a single black
hole, have to be compared with the (single, static, spherically
symmetric) dilaton black--hole solution of Refs.~\cite{kn:dilbh}, which
can be rewritten as follows

\begin{eqnarray}
ds^{2} & = & Vdt^{2}-Wd\vec{x}^{2}\, ,
\nonumber \\
& &
\nonumber \\
V & = & \frac{1}{(A-B)^{2}} \left[ A(\psi/\chi)^{1/2}
-B(\psi/\chi)^{-1/2} \right]^{2}\, ,
\nonumber \\
& &
\nonumber \\
W & = & (\psi\chi)^{2}
\left[ A(\psi/\chi)^{1/2}+B(\psi/\chi)^{-1/2} \right]^{2}\, ,
\nonumber \\
& &
\nonumber \\
e^{-2\phi} & = & e^{-2\phi_{0}}
\left[ A(\psi/\chi)^{1/2}+B(\psi/\chi)^{-1/2} \right]^{-2}\, ,
\nonumber \\
& &
\nonumber \\
F_{\hat{\imath}t} & = & \pm \frac{e^{\phi_{0}}}{\sqrt{2}(A-B)}
\partial_{\hat{\imath}}
\left[ A(\psi/\chi)^{1/2}+B(\psi/\chi)^{-1/2} \right]^{2}\, ,
\end{eqnarray}

\noindent where

\begin{eqnarray}
\psi & = & 1+\frac{M-\Sigma-\sqrt{-4M\Sigma}}{2r}\, ,
\nonumber \\
& &
\nonumber \\
\chi & = & 1+\frac{M-\Sigma+\sqrt{-4M\Sigma}}{2r}\, ,
\end{eqnarray}

\noindent and

\begin{eqnarray}
A+B & = & 1\, ,
\nonumber \\
A-B & = & \sqrt{\frac{-\Sigma}{M}}\, .
\end{eqnarray}

Obviously, the initial data Eq.~(\ref{eq:emdbh2}) for more than one
black hole cannot be compared with any known extact static solution.

By allowing more general values of $q$ in Eq.~(\ref{eq:t}) one gets the
more general family of initial data

\begin{eqnarray}
dl^{2} & = & (\psi/\chi)^{2} \left[ A(\psi/\chi)^{c}
+B(\psi/\chi)^{-c} \right]^{2} d\vec{x}^{2}\, ,
\nonumber \\
& &
\nonumber \\
e^{-2\phi} & = & e^{-2\phi_{0}} \left[ A(\psi/\chi)^{c}
+B(\psi/\chi)^{-c} \right]^{-2}\, ,
\nonumber \\
& &
\nonumber \\
Z & = & C \pm \frac{\sqrt{1-2c^{2}}}{c} e^{\phi_{0}}
\left[ A(\psi/\chi)^{c} -B(\psi/\chi)^{-c} \right]\, ,
\label{eq:emdbh3}
\end{eqnarray}

\noindent where the constant $c$ taks values in the interval
$[-1/\sqrt{2},+1/\sqrt{2}]$.  This family of initial data includes that
of Eqs.~(\ref{eq:emdbh1}) ($c=\pm b$ and $A=1$ or $B=1$) and
Eqs.~(\ref{eq:emdbh2}) ($c=1/2$ and $(A+B)^{2}=1$).  More general
solutions based on the transformation Eq.~(\ref{eq:t}) are possible if
one allows the constant $c$ to be imaginary.  I will not treat them
here, but some of them will be important when we study the wormhole
initial--data problem in Section~\ref{sec-worm}.  Nevertheless, I would
like to make the following observation here: if one performs the duality
transformation Eq.~(\ref{eq:dual}), then the rescaled string--frame
metric $e^{2\phi}dl^{2}$ is the same as the Einstein--Maxwell one
Eq.~(\ref{eq:embh}).

It is easy to show again that in each of the limits $r_{i}\equiv|\vec{x}
-\vec{x}_{i}|\rightarrow 0$ there is another asymptotically flat region.
In this limit the metric looks like this

\begin{eqnarray}
dl^{2} & \sim &
\left[ (1 +\psi^{\prime}_{i}/r^{\prime}_{i})/(1
+\chi^{\prime}_{i}/r^{\prime}_{i}) \right]^{2}
\{A^{\prime}
\left[ (1 +\psi^{\prime}_{i}/r^{\prime}_{i})/(1
+\chi^{\prime}_{i}/r^{\prime}_{i}) \right]^{c}
\nonumber \\
& &
\nonumber \\
& &
+B^{\prime}
\left[ (1 +\psi^{\prime}_{i}/r^{\prime}_{i})/(1
+\chi^{\prime}_{i}/r^{\prime}_{i}) \right]^{-c}\}
\left[ dr_{i}^{\prime 2} +r_{i}^{\prime 2}d\Omega^{2} \right]\, ,
\label{eq:asympemd}
\end{eqnarray}
where
\begin{eqnarray}
r_{i}^{\prime} & = &
\left[ A(\psi_{i}/\chi_{i})^{c} +B(\psi_{i}/\chi_{i})^{-c}
\right]/r_{i}^{-1}\, , \nonumber \\
& &
\nonumber \\
A^{\prime} & = & A(\psi_{i}/\chi_{i})^{c}/
\left[ A(\psi_{i}/\chi_{i})^{c} +B(\psi_{i}/\chi_{i})^{-c} \right]\, ,
\nonumber \\
& &
\nonumber \\
B^{\prime} & = & B(\psi_{i}/\chi_{i})^{-c}/
\left[ A(\psi_{i}/\chi_{i})^{c} +B(\psi_{i}/\chi_{i})^{-c} \right]\, ,
\nonumber \\
& &
\nonumber \\
\psi^{\prime}_{i} & = & \chi_{i}
\left[ A(\psi_{i}/\chi_{i})^{c} +B(\psi_{i}/\chi_{i})^{-c} \right]
\left( 1 +\sum_{j\neq i}^{N} \frac{\psi_{j}}{r_{ij}} \right)\, ,
\nonumber \\
& &
\nonumber \\
\chi^{\prime}_{i} & = & \psi_{i}
\left[ A(\psi_{i}/\chi_{i})^{c} +B(\psi_{i}/\chi_{i})^{-c} \right]
\left( 1 +\sum_{j\neq i}^{N} \frac{\chi_{j}}{r_{ij}} \right)\, ,
\nonumber \\
& &
\nonumber \\
r_{ij} & = & |\vec{x}_{i}-\vec{x}_{j}|\, .
\end{eqnarray}

For the metric in Eq.~(\ref{eq:emdbh3}) to be regular the positivity of
the constants $\psi_{i}$ and $\chi_{i}$ is not sufficient.  One has to
impose certain restrictions on the values of the constants $A$ and $B$
as well.  When both have the same sign, the metric is always regular,
but when they have opposite signs one needs to study in detail the
problem.  I will do it for the case of a single black hole in
Section~\ref{subsec-emdss}.

The conclussion is that for any $N$, if the constants $A$ and $B$ are
carefully chosen, the initial data Eq.~(\ref{eq:emdbh3}) (and hence
those of Eq.~(\ref{eq:emdbh2})) describe several charged dilaton black
holes of the same type at the moment of time--symmetry.

It seems that the initial data of Eqs.~(\ref{eq:emdbh2}) are those
seeked for.  However, the initial data in Eqs.~(\ref{eq:emdbh3}) {\it
also} describe charged dilaton black holes!  And the possibilities of
generating new solutions are not exhausted.  It seems that now there are
too many solutions.  For given physical parameters $Q$, $P$, $M$ and
$\Sigma$, how many different solutions are there now?  Why are they
different?  I will address this problem, already present in the
Einstein--dilaton system, in the next section by studying the simplest
solutions in these families: the spherically symmetric, which describe a
single black hole.


\section{The spherically symmetric case}
\label{sec-sphesymm}

Throughout this section I will take the harmonic functions $\psi,\chi$
to be

\begin{eqnarray}
\psi & = & 1+\frac{E}{r}\, ,
\nonumber \\
& &
\nonumber \\
\chi & = & 1+\frac{F}{r}\, .
\label{eq:ss}
\end{eqnarray}

In most cases, the regularity of the solutions implies that $E$ and $F$
are strictly positive constants and, then, all the initial--data sets
found in the previous section describe a single spherically symmetric
black hole.  In this section I want to identify the physical meaning of
the constants $E$ and $F$ in the different asymptotic regions,
expressing $E$ and $F$ in terms of the mass, and charges.  My goal is to
find out if there is more than one initial--data set describing a
spherically symmetric black hole solutions with fixed mass and charges,
and whether there are any extra degrees of freedom.


\subsection{Einstein--dilaton}

In the upper sheet ($r \rightarrow \infty$), the mass, the dilaton
charge and the asymptotic value of the dilaton are

\begin{eqnarray}
M & = & (1+a)E +(1-a)F\, ,
\nonumber \\
& &
\nonumber \\
\Sigma & = & \pm\sqrt{1-a^{2}} (E-F)\, ,
\nonumber \\
& &
\nonumber \\
\phi_{\infty} & = & \phi_{0}\, .
\end{eqnarray}

Observe that the strict positivity of $E$ and $F$ and the fact that
$|a|<1$ ensure the strict positivity of the mass.

It is convenient to have the expressions of $E$ and $F$ in terms of the
physical constants $M$ and $\Sigma$, and the parameter $a$:

\begin{eqnarray}
E & = & {\textstyle\frac{1}{2}}\biggl (M \pm
\sqrt{\frac{1-a}{1+a}}\Sigma\biggr )\, ,
\nonumber \\
& &
\nonumber \\
F & = & {\textstyle\frac{1}{2}}\biggl (M \mp
\sqrt{\frac{1+a}{1-a}}\Sigma\biggr)\, .
\end{eqnarray}

These constants are positive and, therefore, the mass obeys the bounds

\begin{equation}\label{eq:edbound}
M> \mp \sqrt{\frac{1-a}{1+a}}\, \Sigma\, ,
\hspace{1cm}
M> \pm \sqrt{\frac{1+a}{1-a}}\, \Sigma\, ,
\end{equation}

\noindent which means that, for fixed value of the dilaton charge, there
is always a value of the parameter $a$ such that the mass is as small as
we please and it is only bounded by zero in this family of initial data.

The radius of the minimal surface (there is only one) is given in terms
of the integration constants $E,F,a$ by

\begin{equation}\label{eq:r0ed1}
r_{min} ={\textstyle \frac{1}{2}} \biggl \{ a(E-F) +\sqrt{a^{2}
(E-F)^{2} +4EF} \biggr \} \, ,
\end{equation}

\noindent and, in terms of the physical constants and the paremeter
$a$ by

\begin{equation}\label{eq:r0ed2}
r_{min} ={\textstyle \frac{1}{2}}
\biggl \{ \pm \frac{a}{\sqrt{1 -a^{2}}}\Sigma
+\sqrt{(M \mp\frac{a \Sigma}{\sqrt{1-a^{2}}})^{2}
-\Sigma^{2}} \biggr \} \, .
\end{equation}

It is easy to see in Eq.~(\ref{eq:r0ed1}) that the positivity
of $E$ and $F$ imply that $r_{min}$ is always real and positive, {\it
i.e.} there is always an apparent horizon. From the expression of
$r_{min}$ in terms of the physical constants and the fact the it is real
and positive one sees that the charges obey a number of bounds which
reduce to those in Eq.~(\ref{eq:edbound}).

The area of the minimal surfaces can be readily found, but it is a
complicated and not very enlightening expression. For small dilaton
charge one finds

\begin{equation}
A(r_{min}) =16\pi (M^{2} -\frac{1}{2}\Sigma^{2}) +O(\Sigma^{3})\, ,
\end{equation}

\noindent which does not depend on the parameter $a$ to this order and
coincides with the area of the event horizon of a Reissner--Nordstr\"om
black hole with  electric charge $Q=\Sigma$.

{}From these expressions it is easy to see that this is a one--parameter
family of initial data describing a black hole of fixed mass and dilaton
charge. Which additional degree of freedom is $a$ describing? To answer
this question one has to calculate first the values of the mass and
dilaton charge as seen from the second asymptotic region.

To study the ``lower sheet" one has to perform first the coordinate
transformation $r\rightarrow r^{\prime}=E^{1+a}F^{1-a}/r$. The metric
and the dilaton field are in these new coordinates

\begin{eqnarray}
dl^{2} & = &
\left( 1 +\frac{E^{a} F^{1-a}}{r^{\prime}} \right)^{2+2a}
\left( 1 +\frac{E^{1+a} F^{-a}}{r^{\prime}} \right)^{2-2a}
(dr^{\prime 2} +r^{\prime 2} d\Omega^{2})\, ,
\nonumber \\
& &
\nonumber \\
\phi & = & \phi_{0} \pm \sqrt{1-a^{2}} \log(E/F)
\nonumber \\
& &
\nonumber \\
& &
\pm\log \left[
\left( 1 +\frac{E^{a} F^{1-a}}{r^{\prime}} \right)/
\left( 1 +\frac{E^{1+a} F^{-a}}{r^{\prime}} \right) \right]\, .
\end{eqnarray}

\noindent One immediately gets in the limit $r^{\prime} \rightarrow
\infty$

\begin{eqnarray}
M^{\prime} & = & (1+a)E^{a}F^{1-a} +(1-a)E^{1+a}F^{-a}\, ,
\nonumber \\
& &
\nonumber \\
\Sigma^{\prime} & = & \pm \sqrt{1-a^{2}} (E^{a}F^{1-a}
-E^{1+a}F^{-a})=-(E/F)^{a}\Sigma\, ,
\nonumber \\
& &
\nonumber \\
\phi_{\infty}^{\prime} & = & \phi_{0} \pm\sqrt{1+a^{2}}\log(E/F)\, .
\label{eq:chlw}
\end{eqnarray}

When $a\neq 0$ all the physical constants are different in the upper and
lower sheets.  These results may seem surprising at first (they are
certainly unusual) but, for $M$ and $\Sigma$, they are simple
consequences of the absence of a Gauss' law for the dilaton charge, as I
am going to explain.

In the Reissner--Nordstr\"om initial data problem the Gauss' law says
that there is no charge on the initial surface $\Sigma$.  The nontrivial
topology of $\Sigma$ allows the electric force lines to get {\it
trapped} and go from the upper sheet to the lower sheet across the
Einstein--Rosen bridge.  The electric flux through an asymptotic sphere
in each separate sheet does not vanish, but the total flux does, in
agreement with the Gauss' law.  The effect is that observers in both
sheets measure the same amount of charge but with opposite signs.  I
will call this type of charge {\it topological charge}.

In the present case there is no Gauss' law for $\phi$.  The presence of
certain amount of {\it physical} scalar charge $\Sigma_{p}$ on the
initial surface is allowed. It corresponds to the charge density

\begin{equation}
\rho={}^{(3)}\nabla^{2}\phi
=\partial_{\hat{\imath}}(W^{\frac{1}{2}}\partial_{\hat{\imath}}\phi)\, .
\end{equation}

Integrating $\rho$ on $\Sigma$ and applying Gauss' theorem

\begin{equation}
\frac{1}{4\pi}\int_{\Sigma}d^{3}x \rho=
\frac{1}{4\pi} \int_{S^{2}_{up}}
dS^{\hat{\imath}} {}^{(3)}\nabla_{\hat{\imath}}\phi+
\frac{1}{4\pi} \int_{S^{2}_{low}}
dS^{\hat{\imath}} {}^{(3)}\nabla_{\hat{\imath}}\phi
\equiv 2\Sigma_{p}\, ,
\end{equation}

\noindent where $S^{2}_{up}$ and $S^{2}_{low}$ are two two--spheres in
the limits $r \rightarrow \infty$ and $r \rightarrow 0$ respectively.

If there was only this charge, one would measure exactly the same charge
$\Sigma_{p}$ in both sheets, (with the same signs).  As one can see in
Eq.~(\ref{eq:chlw}), this is not the case, in general.  The force lines
of the field $\phi$ can also get trapped in the throat\footnote{It
should be stressed that this is simply a good way of describing the
system.  There is no conserved charge of any kind associated to the
dilaton.  I will keep on using the terms ``charge" and ``topological"
charge although it should be clear that in this system there is no
conserved dilaton charge of any kind.}.  This effect contributes with
different signs to the flux of $\phi$ in both sheets: $\pm\Sigma_{t}$.
The result is that the flux of $\phi$ in the upper sheet is $\Sigma
=\Sigma_{p} +\Sigma_{t}$ and in the lower sheet is $\Sigma^{\prime}
=\Sigma_{p} -\Sigma_{t}$:

\begin{eqnarray}
\Sigma & = & \frac{1}{4\pi} \int_{S^{2}_{up}} dS^{\hat{\imath}}
{}^{(3)}\nabla_{\hat{\imath}}\phi  =\Sigma_{p} +\Sigma_{t}\, ,
\nonumber \\
& &
\nonumber \\
\Sigma^{\prime} & = & \frac{1}{4\pi} \int_{S^{2}_{low}}
dS^{\hat{\imath}}
{}^{(3)}\nabla_{\hat{\imath}}\phi =\Sigma_{p}-\Sigma_{t}\, .
\end{eqnarray}

This explains the physical meaning of the parameter $a$ as well: it
measures the relative value of $\Sigma_{p}$ versus $\Sigma_{t}$ which
was the physical degree of freedom that we failed to identify before:

\begin{equation}
\Sigma_{p}/\Sigma = {\textstyle \frac{1}{2}}(\Sigma
+\Sigma^{\prime})/\Sigma ={\textstyle \frac{1}{2}} \left[ 1
-(E/F)^{a} \right] <\frac{1}{2}\, .
\end{equation}

$\Sigma_{p}$ is always different from $\Sigma$ except when $\Sigma=0$
{\it i.e.} there is always some topological and some physical dilaton
charge except in two cases: when there is no dilaton charge at all and
when $a=0$, the Reissner--Nordstr\"om--like case in which all the charge
is ``topological".

This justifies the difference between $\Sigma$ and $\Sigma^{\prime}$.
The difference between $M$ and $M^{\prime}$ must correspond to the
difference in the matter contents (scalar charge) observed in both
regions.

Finally, observe that most of these effects disappear if one rescales
the metric by $\exp\{\frac{-2a}{\sqrt{1-a^2}}\phi\}$: the mass and
dilaton charge in the upper and lower sheets are the same, but the
asymptotic value of the dilaton, $\phi_{\infty}$ is still different.


\subsection{The Einstein--Maxwell--dilaton case}
\label{subsec-emdss}

As it was pointed out in Section~\ref{subsec-emd}, the regularity of the
metric in Eqs.~(\ref{eq:emdbh3}), (\ref{eq:ss}) is not guaranteed by the
simple positivity of the constants $E$ and $F$, and additional
conditions have to be imposed to the constants $A$ and $B$.  The first
thing to do in order to study these initial data is to find these
conditions.

In what follows I will assume that $A>0$ with no loss of generality.
If $B>0$ too, then the metric is obviously regular. If $B<0$ the metric
vanishes when the coordinate $r$ takes the value

\begin{equation}
r_{sing}=\frac{E -(-B/A)^{1/2c}}{(-B/A)^{1/2c} -1}\, ,
\end{equation}

\noindent and therefore it will be regular if $r_{sing}<0$.  A careful
analysis leads to the identification of the following five cases in
which the metric is regular

\begin{enumerate}

\item $B\geq 0$, $A+B=+1$,

\item $B<0$, $A+B=+1$, $(E/F)^{2c}>1$,

\item $B<0$, $A+B=+1$, $(E/F)^{2c}<1$,  $A<1/[1 -(E/F)^{2c}]$,

\item $B<0$, $A+B=-1$, $(E/F)^{2c}>1$, $A<-1/[1 -(E/F)^{2c}]$,

\item $B<0$, $A+B=-1$, $(E/F)^{2c}<1$,

\end{enumerate}

In principle in this five instances the positivity of the mass in the
upper and lower sheets should be automatically guaranteed, but a
detailed study of each case is required to prove it.

In the upper sheet the mass, the dilaton charge, the electric charge
and the asymptotic value of the dilaton are
\begin{eqnarray}
M & = & \left[1 +c\frac{(A-B)}{(A+B)} \right] E
+\left[1 -c\frac{(A-B)}{(A+B)} \right] F\, ,
\nonumber \\
& &
\nonumber \\
\Sigma & = & c\frac{(A-B)}{(A+B)}(E-F)\, ,
\nonumber \\
& &
\nonumber \\
Q & = & \pm (A+B) e^{\phi_{0}} \sqrt{1-2c^{2}} (E -F)\, ,
\nonumber \\
& &
\nonumber \\
\phi_{\infty} & = & \phi_{0}\, .
\end{eqnarray}

In the first case above it is easy to prove that $M>0$.  One simply has
to observe that $A>0$, $B\geq 0$, $A+B=1$ imply $|(A-B)/(A+B)|\leq 1$.
This, together with $|c|\leq 1/\sqrt{2}$ imply $|c(A-B)/(A+B)| \leq
1/\sqrt{2}<1$ and the factors that multiply $E$ and $F$ in the mass
formula are strictly positive.  In the second case it is also easy to
prove that $M>0$ by rewritting the mass formula as follows:

\begin{equation}
M=(E+F)+c\frac{(A-B)}{(A+B)}(E-F)\, .
\end{equation}

\noindent Now the product $c(E-F)>0$ in this case and all the terms in
the above expression are manifestly positive.

The third case is more complicated. One has to study two distinct
possibilities: (i) $E>F$, $c<0$ and (ii) $E<F$, $c>0$.
One has to prove that $A<1/[1 -(E/F)^{2c}]$ implies $A<\frac{1}{2}
\left[1 -\frac{1}{c} \left( \frac{E+F}{E-F} \right) \right]$.
In the case (i), if $c<-1/2$, it follows from the inequality

\begin{equation}
\frac{1}{1-(F/E)^{-2c}}<\frac{1}{1-(F/E)}\, ,
\end{equation}

\noindent and, if $c>-1/2$ it follows from

\begin{equation}
\frac{1}{1-(F/E)^{-2c}}<\frac{-1}{2c} \left[ \frac{1}{1-(F/E)} \right]
\, .
\end{equation}

The case (ii) follows from (i) with the interchange $E\leftrightarrow F$
$c\rightarrow -c$.  The remaining two cases go through in a similar
fashion and I will omit the details.

These mass and charges formulae can be inverted to express the
integration constants in terms of the physical constants:
\begin{eqnarray}
E & = & {\textstyle\frac{1}{2}} \biggl (M -\Sigma
\pm\frac{e^{-\phi_{\infty}}}{\sqrt{1-2c^{2}}} Q\biggr )\, ,
\nonumber \\
& &
\nonumber \\
F & = & {\textstyle\frac{1}{2}}\biggl (M -\Sigma
\mp\frac{e^{-\phi_{\infty}}}{\sqrt{1-2c^{2}}} Q\biggr )\, ,
\nonumber \\
& &
\nonumber \\
c(A-B) & = & \pm \sqrt{1-2c^{2}} e^{-\phi_{\infty}} \frac{\Sigma}{Q}\, .
\end{eqnarray}

Again the number of independent integration constants is the same as the
number of physical constants plus one.  It is clear that there is again
another degree of freedom described by the extra integration constant.
If one calculates now the value of $M,\Sigma,Q,\phi_{\infty}$ in the
lower sheet one would find different values for the four of them.  This
seems to suggest that there is one less integration constant than
necessary.  This confussion arises because the ``electric" charge to
look for is $\tilde{Q} =e^{-2\phi_{\infty}} Q$ for which there is a
Gauss' law ${}^{(3)}\nabla_{\hat{\imath}} (e^{-2\phi} E^{\hat{\imath}})
=0$ enforcing its absence on the initial surface.  This equation implies
that all the $\tilde{Q}$ charge is ``topological" and that it takes
equal values with different signs in the upper and lower sheets in
agreement with the arguments of the previous section.  This can be
readily checked.

The integration constant $c$ measures the ratio between topological and
physical dilaton charge for which there is no Gauss' law whatsoever.

The radii $r_{min}$ of the minimal surfaces are given by the zeros of
the funcion

\begin{equation}
f(r) =A (r+E)^{2c} (r-r_{+}) (r-r_{-}) +B (r+F)^{2c} (r+r_{+})
(r+r_{-})\, ,
\end{equation}

\noindent where

\begin{equation}
r_{\pm}= \frac{1}{2} \left[ -c(F-E) \pm\sqrt{c^{2}(F-E)^{2} +4EF}
\right]\, .
\end{equation}

Finding an analytical expression for $r_{min}$ is out of the question.
Nevertheless, observing that the first term of $f(r)$ only vanishes
at $r=r_{+}$ and the second term only vanishes at $r=r_{-}$ it is
possible to say the following about it:

\begin{enumerate}

\item If $sign(A)=sign(B)$, $c(F-E)>0$, then $r_{min} \in [r_{+},
-r_{-}]$.

\item If $sign(A)=sign(B)$, $c(F-E)<0$, then $r_{min} \in [-r_{-},
r_{+}]$.

\item If $sign(A)=-sign(B)$, $c(F-E)>0$, then $r_{min} \in
(0,r_{+}]\cup [-r_{-},\infty)$.

\item If $sign(A)=-sign(B)$, $c(F-E)>0$, then $r_{min} \in
(0,-r_{-}]\cup [r_{+},\infty)$.

\end{enumerate}


\section{Wormhole initial data}
\label{sec-worm}

To find wormhole initial data I will follow Misner Ref.~\cite{kn:Mis}.
I will take the $\xi$--metric to be the metric of the ``doughnut"
$S^{1}\times S^{2}$ Eq.~(\ref{eq:doughnut}). The effect of the conformal
factor $W$ on this metric is to blow up one side of the doughnut and
transform it in an asymptotically flat region. This metric has constant
curvature ${}^{(3)}R_{D}=2$. Then, the kind of equation one has to solve
for all the ansatzs of Section~\ref{sec-ansatzs} is

\begin{equation}\label{eq:eso}
(\nabla_{D}^{2} -{\textstyle\frac{1}{4}})\chi = 0\, .
\end{equation}

Misner observed that, although this equation looks complicated, there
is one simple solution. In fact, the metric of flat three--space in
bispherical coordinates is

\begin{equation}
dl^{2}_{\flat}=\frac{k^{2}}{(\cosh\mu-\cos\theta)^{2}}dl^{2}_{D}\, ,
\end{equation}

\noindent and must be a solution of the initial--data constraint
equations in vacuum. Therefore the functions

\begin{equation}
f_{d}(\theta,\mu)=k^{-1/2}[\cosh(\mu+d)-\cos\theta]^{-1/2}\, ,
\end{equation}

\noindent where $d$ is any constant must solve Eq.~(\ref{eq:eso}).
Now, of course, this fact can be used not just for the vacuum case but
for all the cases in which the $\xi$--metric is the doughnut metric of
Eq.~(\ref{eq:doughnut}) and one has been able to reduce the initial--data
problem to equations like Eq.~(\ref{eq:eso}).  One just has to build
$\chi$ and $\psi$ as linear combinations of functions $f_{d}$ obeying
the boundary conditions required in each case.

Essentially the boundary conditions can be described as follows: In the
wormhole space the coordinate $\mu$ crosses the wormhole and
parametrizes the $S^{1}$.  Therefore $\mu$ is periodic with
period\footnote{Observe that our convention differs slightly from that
of Ref.~\cite{kn:Mis} and the subsequent litarature in which $\mu$ has
period $2\mu_{0}$.} $\mu_{0}$.  The metric, the electric field and the
dilaton (which are physical fields) must be single--valued around the
wormhole.  That is, all those fields must be periodic in the variable
$\mu$ with period $\mu_{0}$.  The electrostatic potential is not
physical and can be multivalued around the wormhole.

To construct these periodic fields it is convenient to define first the
functions $f_{n}(\theta,\mu)=f_{0}(\theta,\mu+n\mu_{0})$ where $n$ is an
integer number.  The series

\begin{equation}\label{eq:series}
\sum_{n\in Z}c_{n}f_{n}(\theta,\mu)\, ,
\end{equation}

\noindent is a solution of the linear Equation~(\ref{eq:eso}) with
different periodicity properties depending on the choice of the
constants $c_{n}$.  If one chooses all the $c_{n}=1$, the series is
periodic with period $\mu_{0}$ (assuming it converges).  For
$c_{n}=\pm(-1)^{n}$ the series is antiperiodic.  Many more choices are
possible, leading to different periodicity properties, and some of them
will be used later.  Now I will study each case separately.


\subsection{Vacuum (Misner) wormholes}

The single--valuedness of $W=\chi^{4}$ implies $c_{n}\sim e^{\frac{i\pi
n }{2}}$.  Reality and regularity of $W$ imply $c_{n}=1$ for all
integers $n$, and so

\begin{equation}
\chi=k^{-1/2}\sum_{n\in Z}[\cosh(\mu+n\mu_{0})-\cos\theta]^{-1/2}\, .
\end{equation}

\noindent which is the solution found by Misner in Ref.~\cite{kn:Mis}.
For a external observer the wormhole's two mouths are just two
Schwarzschild black holes.  The fact that there are no more
asymptotically flat regions (``universes") as in the initial data found
in the previous Section seems untestable for that observer since the
wormhole's throat (or the Einstein--Rosen bridge between universes)
collapses before any signal crosses it.  These initial data were used by
Smarr in Ref.~\cite{kn:Smarr} to study the gravitational radiation
produced in a head--on collision of two Schwarzschild black holes.  The
evolution of these initial data have also been studied in a different
limit by Tomimatsu in Ref.~\cite{kn:Tom}.


\subsection{Einstein--Maxwell wormholes}

The single--valuedness of $W$ in Eq.~(\ref{eq:emans}) requires that, if
the functions $\chi$ and $\psi$ get a factor $K$ when moving from $\mu$
to $\mu+\mu_{0}$, then

\begin{eqnarray}
\psi(\theta,\mu+\mu_{0}) & = & K\psi\ (\theta,\mu)\, ,
\nonumber \\
& &
\nonumber \\
\chi(\theta,\mu+\mu_{0}) & = & K^{-1}\chi(\theta,\mu)\, .
\end{eqnarray}

\noindent Regularity and reality of $W$ require $K$ to be a positive
real number wich is customarily written as $K=e^{-\lambda}$.  When
$\lambda=0$ one is back into the vacuum case.

Functions with the required monodromy properties can be built as the
series Eq.~(\ref{eq:series}) with coefficients $c_{n}=e^{n\lambda}$ and
$c_{n}=e^{-n\lambda}$ respectively, that is

\begin{eqnarray}
\psi & = & k^{1/2}\sum_{n\in Z} e^{n\lambda}[\cosh
\mu-\cos\theta]^{-1/2}\, ,
\nonumber \\
& &
\nonumber \\
\chi & = & k^{1/2}\sum_{n\in Z} e^{-n\lambda}[\cosh
\mu-\cos\theta]^{-1/2}\, .
\label{eq:emw}
\end{eqnarray}

Observe that with this choice of monodromy of $\chi$ and $\psi$ around
$\mu$, the electrostatic potential $Z$ is not single--valued

\begin{equation}
Z(\theta,\mu+\mu_{0})=Z(\theta,\mu)\mp 2\lambda\, ,
\label{eq:multipot}
\end{equation}

\noindent but the physically meaningful quantity, the electric field,
is single--valued.  These are the electrically charged wormhole initial
data found by Lindquist by the method of images in Ref.~\cite{kn:Lin}.


\subsection{Einstein--dilaton wormholes}

The single--valuedness of the metric in Eq.~(\ref{eq:emdans}) requires
now

\begin{eqnarray}
\psi(\theta,\mu+\mu_{0}) & = & K^{\frac{1}{2(1+a)}}\psi(\theta,\mu)\, ,
\nonumber \\
& &
\nonumber \\
\chi(\theta,\mu+\mu_{0}) & = & K^{\frac{-1}{2(1-a)}}\chi(\theta,\mu)\, .
\end{eqnarray}

I will write $K=e^{-2\lambda}$ to recover the previous case when $a=0$.
The functions $\psi$ and $\chi$ are the series Eq.~(\ref{eq:series})
with coefficients $c_{n}=e^{\frac{n\lambda}{1+a}}$ and
$c_{n}=e^{\frac{-n\lambda}{1-a}}$ respectively

\begin{eqnarray}
\psi & = & k^{1/2}\sum_{n\in Z} e^{\frac{n}{1+a}\lambda}[\cosh
(\mu+n\mu_{0})-\cos\theta]^{-1/2}\, ,
\nonumber \\
& &
\nonumber \\
\chi & = & k^{1/2}\sum_{n\in Z} e^{\frac{-n}{1-a}\lambda}[\cosh
(\mu+n\mu_{0})-\cos\theta]^{-1/2}\, .
\end{eqnarray}

Now, analyzing the monodromy properties of the dilaton field, one finds
that

\begin{equation}
\phi(\theta,\mu+\mu_{0})=\phi(\theta,\mu)
\mp\frac{2\lambda}{\sqrt{1+a^{2}}}\, ,
\end{equation}

\noindent that is, the dilaton is not single--valued around the wormhole,
but its value changes by a constant each time one goes around the the
wormhole, just as it happened with the electrostatic potential.  Since
the zero mode of the dilaton is physically meaningful, one conclude that
this is not a good solution of the initial data problem for a dilaton
field such as the string theory one.

One could have anticipated this result because, roughly speaking, to
build a wormhole one has to identify the two asyptotic regions of an
Einstein--Rosen bridge and in Section~\ref{sec-sphesymm} we found that
the asymptotic value of the dilaton in both regions is, in general,
different and cannot be identified.

Was one considering a different kind of field, for instance an scalar
taking values on a circle whose length is a submultiple of
$\frac{2\lambda}{\sqrt{1+a^{2}}}$ the solution would be perfectly valid.
If the length of this circle is $\frac{2\lambda}{n\sqrt{1+a^{2}}}$, $n
\in Z^{+}$, then one is identifying $\phi$ with $\phi +m
\frac{2\lambda}{n\sqrt{1+a^{2}}}$, for all integers $m$.  This scalar
field is a map from one circle parametrized by $\mu$ to another circle
(the target), such that going around the first circle once means going
around the target circle $\pm n$ times, and one can consider this number
as the winding number of the map.  This is a topological invariant of
$\phi$ that cannot change if the initial surface $\Sigma$ is deformed in
a continuous fashion.  Therefore, as long as $\Sigma$ does not become
singular, the winding number of the field configuration $\phi$ will not
change in the time--evolution of the initial data.

Another case would be that of an axion field $a$.  One might consider
that what has physical meaning is not $a$ but $\partial a$.  The zero
mode being meaningless, the solution would be valid as well.


\subsection{Einstein--Maxwell--dilaton wormholes}

The single--valuedness of the metric in our ansatz Eq.~(\ref{eq:emdans})
requires

\begin{eqnarray}
\psi(\theta,\mu+\mu_{0}) & = & K^{\frac{1}{2(1+b)}}\psi(\theta,\mu)\, ,
\nonumber \\
& &
\nonumber \\
\chi(\theta,\mu+\mu_{0}) & = & K^{\frac{-1}{2(1-b)}}\chi(\theta,\mu)\, .
\end{eqnarray}
Writing again $K=e^{-2\lambda}$ we get
\begin{eqnarray}
\psi & = & k^{1/2}\sum_{n\in Z} e^{\frac{n}{1+b}\lambda}[\cosh
(\mu+n\mu_{0})-\cos\theta]^{-1/2}\, ,
\nonumber \\
& &
\nonumber \\
\chi & = & k^{1/2}\sum_{n\in Z} e^{\frac{-n}{1-b}\lambda}[\cosh
(\mu+n\mu_{0})-\cos\theta]^{-1/2}\, .
\end{eqnarray}

Now let us examine the monodromy properties of the dilaton and the
electrostatic potential. Going around the wormhole once

\begin{eqnarray}
\phi(\theta,\mu+\mu_{0}) & = & \phi(\theta,\mu)
\mp\frac{2\lambda}{\sqrt{1+b^{2}}}\, ,
\nonumber \\
& &
\nonumber \\
Z(\theta,\mu+\mu_{0}) & = &
e^{\frac{-2b\lambda}{1-b^{2}}}Z(\theta,\mu)\, .
\end{eqnarray}

\noindent The electrostatic potential changes by a factor, and so does
the electric field.  This means that this is not a valid solution even
accepting the multivaluedness of the dilaton field.

Fortunately there is another procedure to get a wormhole solution for
the Einstein--Maxwell--dilaton system.  As I showed in
Section~\ref{subsec-emd} there is a way of mapping initial data of the
Einstein--Maxwell system into initial data of the
Einstein--Maxwell--dilaton system.  One could start from the
Reissner--Nordstr\"om wormhole Eqs.~(\ref{eq:embh}),(\ref{eq:emw}) and
transform it by means of a function $t(Z)$. According to
Eq.~(\ref{eq:multipot}), the electrostatic
potential $Z$ of the Reissner--Nordstr\"om wormhole is not a
single--valued
function in the initial surface. Its value
changes by an amount of $T=\pm 2\lambda$.  Since $t(Z)$ is the new
electrostatic potential it is clear that one simply needs a function
$t(Z)$ periodic in $Z$ with a period which is a submultiple of $T$ and
satisfies the differential Equation~(\ref{eq:differential}).  This would
also guarantee the single--valuedness of the dilaton field which is a
power of $dt/dZ$.

The surprising thing is that the function needed for this problem was
already found in Section~\ref{subsec-emd}.  It is given by
Eq.~(\ref{eq:t}) with the constant $q$ given by

\begin{equation}
q=1+\frac{32\pi^{2}\lambda}{n^{2}}\, ,\hspace{1cm}n=1,2,\ldots\, ,
\end{equation}

\noindent and appropriate choices of the constants $A,B,D$, so $t(Z)$ is
simply

\begin{eqnarray}
t(Z) & = & M \sin (\omega Z+\gamma_{0})\, ,
\nonumber \\
& &
\nonumber \\
\omega & = & \frac{4\pi\lambda}{n}\, .
\end{eqnarray}

I rewrite below for convenience the whole wormhole initial data family
for the Einstein--Maxwell--dilaton system:
\begin{eqnarray}
dl^{2} & = & \tilde{W}(d\mu^{2}+d\theta^{2}+ \sin^{2}\theta
d\phi^{2})\, ,
\nonumber \\
& &
\nonumber \\
\tilde{W} & = & p M^{2}\omega^{2}
\cos^{2}[\omega\log (\psi/\chi)+\gamma_{0}](\psi\chi)^{2}\, ,
\nonumber \\
& &
\nonumber \\
e^{-2\tilde{\phi}} & = & \frac{1 +2\omega^{2}}{M^{2}\omega^{2}}
\cos^{-2}[\omega\log (\psi/\chi)+\gamma_{0}]\, ,
\nonumber \\
& &
\nonumber \\
\tilde{Z} & = & M
\sin[\omega\log (\psi/\chi)+\gamma_{0}](\psi\chi)^{2}\, ,
\nonumber \\
& &
\nonumber \\
\psi & = & k^{1/2}\sum_{n\in Z} e^{n\lambda}[\cosh
\mu-\cos\theta]^{-1/2}\, ,
\nonumber \\
& &
\nonumber \\
\chi & = & k^{1/2}\sum_{n\in Z} e^{-n\lambda}[\cosh
\mu-\cos\theta]^{-1/2}\, .
\nonumber \\
& &
\nonumber \\
\omega & = & \frac{4\pi\lambda}{n}\, .
\end{eqnarray}

Although this initial--data set has the required monodromy properties, it
is far from being a regular initial--data set. The metric function
$\tilde{W}$ vanishes in many places, and in those places the dilaton
field blows up.

Observe that, again, after performing an electric--magnetic duality
transformation and shifting the dilaton by an appropriate constant, the
string--frame metric is just that of the Reissner--Nordstr\"om wormhole
Eqs.~(\ref{eq:embh}),(\ref{eq:emw}), perfectly regular.  The dilaton
field $\phi$ still blows up in many places.  This, or so it seems, is
the price one has had to pay for being able to find a single--valued
dilaton field in a space with wormhole topology.


\section{Conclussions}
\label{sec-conclussions}

In this paper I have found several families of time--symmetric
initial--data sets for theories with a masless scalar (dilaton) which
takes values in $R$ or in a circle $S^{1}$.  These families depend on a
certain number of parameters.  For certain values of the parameters,
these solutions describe several black holes (Einstein--Rosen--like
bridges connecting different asymptotically flat regions) in the instant
in which they ``bounce".  Some solutions describe two black holes
connected by a ``wormhole".

In the case of a single black hole it is possible to prove analytically
that the same choice of values of the parameters ensure the regularity
of the solution, the positivity of the mass in the two asymptotically
flat regions and the existence of an apparent horizon.

The presence of a scalar field has many interesting effects.  Perhaps
the most unusual one is that it is more difficult to ``build" initial
surfaces with non--trivial topologies on which the initial data are
regular.  If one had a gauge field on that surface it would be easier to
find solutions in different topologically trivial patches and then glue
them together because the gauge fields of two overlapping patches do not
have to match exactly in the overlap: they only have to match {\it up to
a gauge transformation}.  In the dilaton case there is no gauge
invariance available and the solutions have to match exactly.

Most fields usually considered are not scalars and have gauge
invariances. This is true for the metric, vector fields and axion
two--form (and higher order differential forms). It seems that in
theories containing this kind of fields there are more possible
classical configurations than in theories containing scalars.

In some sense a scalar seems to play the role of a {\it topological
censor}. Of course, more work is necessary to determine to which extent
this is so and which kind of topologies are not allowed if the absence
of singularities and single--valuedness are required.

These results can be generalized and extended to more complex cases:
theories with many scalars (non--linear $\sigma$--models), with scalar
potential, scalar masses etc.  Also, time--symmetric initial--data sets
for other interesting systems besides black holes can be studied.
Particularly interesting in this context are black strings and black
membranes.

Finally, these initial--data sets can be used as the starting point for
investigations on cosmic censorship along the lines of
Ref.~\cite{kn:Gibb} and, perhaps, critical behavior in the gravitational
collapse of a scalar field\footnote{Observe that, at the critical
value of the parameter that measures the energy of the imploding scalar
wave,  a ``point--like" black hole is formed and explodes
again. It is reasonable to expect that the instant of time at which the
black hole exists is a moment of time--symmetry.}. Work on these issues
is in progress.


\section*{Acknowledgements}

The author is indebted to G.W.  Gibbons for attracting his attention to
this problem and for most fruitful discussions and comments.  He would
also like to thank M.M. Fern\'andez for her enduring support.  This work
has been supported by an European Union grant of the {\it Human Capital
and Mobility} program.


\end{document}